\documentclass[10pt, conference, letterpaper]{IEEEtran}

\usepackage{cite}
\usepackage{amsmath,amssymb}
\usepackage{algorithm}
\usepackage{algorithmic}
\usepackage{graphicx}
\usepackage[table]{xcolor}
\usepackage{booktabs}
\usepackage{url}
\definecolor{bandmasrow}{RGB}{255,244,194}


\newcommand{\bandmasrowfill}[1]{\rlap{\raisebox{-\dp\strutbox}{\color{bandmasrow}\rule{#1}{\dimexpr\ht\strutbox+\dp\strutbox\relax}}}}

\begin{document}

\title{BANDMAS: Causality-Inspired Semantic Packet\\
Scheduling for Bandwidth-Efficient\\
Multi-Agent Collaboration}

\author{
    \IEEEauthorblockN{Jiangwen Dong, Wanyu Lin\textsuperscript{*}}
    \IEEEauthorblockA{\textit{Department of Computing, The Hong Kong Polytechnic University}, Hong Kong SAR, China \\
    jiangwen.dong@connect.polyu.hk; wan-yu.lin@polyu.edu.hk; \textsuperscript{*}Corresponding author}
}

\maketitle

\begin{abstract}
LLM-based multi-agent systems make decisions based on the aggregated information via exchanging messages across specialized agents. Forwarding every generated message among agents increases application-layer traffic. Yet, it introduces tremendous input tokens for agent processing, potentially raising inference latency and computational overhead. 
Existing approaches attempt to address the above issues by pruning agents or discarding redundant messages.
Nevertheless, such agent-level or message-level optimization results in insufficient evidence supporting for final decisions or still containing redundant message transmissions.  
To address these challenges, we propose BANDMAS, a multi-agent collaboration framework that models inter-agent communications as task-oriented traffic, which enables efficient transmission via causality-inspired replay valuation. Specifically, we decompose messages into several data packets by analyzing their semantic features such as evidence and requests. The system only transmits these packets if their predicted replay-derived contribution exceeds their resource cost. Consequently, BANDMAS is able to adaptively schedule communication packets while adhering to bandwidth, latency, deadline, and receiver context constraints. 
On frozen Qwen3-4B traffic across SciFact, HotpotQA, and FanOutQA, our framework reduces application-layer bytes by 53.2\% to 77.3\% at selected caps and attains the highest mean task metric among constrained methods on all three workloads. 
Code: \texttt{\url{https://anonymous.4open.science/r/BANDMAS}}
\end{abstract}


\section{Introduction}

LLM-based multi-agent systems divide complex tasks among role-specialized agents that exchange natural-language messages~\cite{wu2023autogen,hong2023metagpt}. In evidence-intensive workflows, reader or specialist agents examine different documents, traces, or telemetry streams, while a downstream verifier or answerer combines their findings. In networked deployments, these agents may run across edge and cloud resources, share an inference service, or belong to different administrative domains. Their intermediate messages must cross application-layer communication paths before reaching a receiver agent for downstream answer generation. This exchange consumes network capacity, expands the input context of the receiver, and increases model-serving computation. Communication therefore determines both the evidence available for the final decision and the resources required to produce it.

The central challenge is that transmitting more message content does not necessarily improve the downstream performance. A single transmitted message may include a claim, supporting evidence, repeated explanations, and requests. However, only partial of this information may affect the final decision. Sending a complete message could waste network and receiver resources, whereas dropping it entirely could lose the necessary evidence. The {\em message length} measures how much content is transmitted, and the {\em semantic similarity} measures how closely that content matches the task; neither establishes whether the transmitted content improves task utility. The central difficulty is to distinguish useful content from the redundant one before the receiver generates an outcome, while preserving the supporting evidence and the dependencies between related claims.

Existing methods typically operate at different stages of the workflow. Agent and edge pruning determine which participants or communication links remain active~\cite{zhang2024agentprune,wang2025agentdropout,safesieve2025,zhang2024sparsedebate}, while workflow routers select an agent composition or topology for each task~\cite{zhang2025gdesigner,chen2025masrouter}. Prompt-compression methods shorten an assembled receiver context~\cite{jiang2023llmlingua,jiang2024longllmlingua,pan2024llmlingua2}, and latent-state systems reduce model-internal transfer or repeated prefill~\cite{ye2025kvcomm,qkvcomm2025}. These methods do not directly determine which parts of a generated message should be transmitted as the receiver agent input. Agent pruning or message pruning methods are too coarse for this purpose, whereas token-level compression may obscure provenance and semantic dependencies. The unresolved problem is to assign task value to interpretable message segments and admit them under wire, deadline, and receiver constraints before observing the receiver agent's eventual outcome.

\begin{figure}[t]
\centering
\includegraphics[width=\columnwidth]{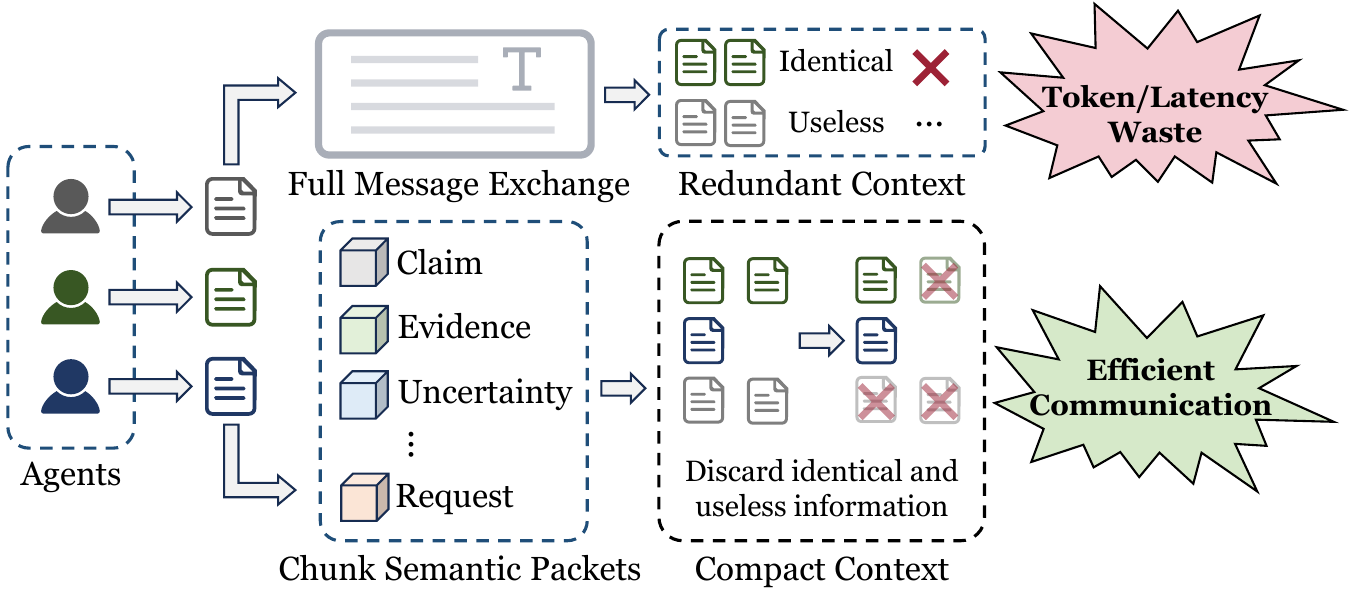}
\caption{Communication-control boundary. Message-level workflows forward complete generated messages. BANDMAS intervenes after generation and before receiver inference, converting each message into traceable semantic packets and admitting complete serialized records under transmission and receiver constraints.}
\label{fig:paradigm-comparison}
\end{figure}

Our intuition is to treat generated inter-agent text as application traffic whose contents have different task values, rather than as an indivisible conversation history. Based on this view, we propose BANDMAS, a receiver-side communication-control framework that packetizes and schedules already-generated content. A \emph{semantic packet} is an application-layer record that carries one message segment and the metadata required for routing, reconstruction, and audit. It is not a physical network packet. This definition separates the communicated content from the schedulable record used for admission and resource accounting. As illustrated in Fig.~\ref{fig:paradigm-comparison}, BANDMAS decides which complete packet records cross the receiver boundary; it does not decide whether an agent may participate or rewrite generated content.

Within this boundary, BANDMAS converts each generated message into traceable semantic packets. This representation provides more granular admission choices than agent or message pruning while preserving the provenance and integrity of each admitted record. BANDMAS then uses a predictor trained on controlled offline replays to estimate each packet's downstream contribution, providing a task-grounded signal that message length or semantic similarity does not directly measure. The scheduler combines these estimates with current communication and processing conditions to determine which complete packets to admit, allowing admission to respond to resource pressure rather than follow a fixed content threshold. While replay is used to learn packet values offline, online scheduling relies only on information available before the final output and never observes gold answers. Together, these techniques enable communication control without changing agent roles, message generation, or the receiver agent.

Overall, this paper makes three contributions:

\begin{itemize}
    \item We formulate LLM-agent communication as resource-constrained semantic packet scheduling, with explicit provenance, dependencies, deadlines, application-layer wire cost, and receiver-context cost.
    \item We design and implement a causality-inspired, split-safe communication-control pipeline that integrates dual-context sufficiency--necessity replay profiling, pre-outcome packet-value prediction, exact wire-cap admission, and resource-price updates using completed transfer and service measurements.
    \item We evaluate BANDMAS across three evidence-aggregation benchmarks using frozen traffic and protocol-matched reproduced baselines, with a two-benchmark Qwen2.5-7B transfer check. At the predeclared FanOutQA 0.50 cap, causal admission has a task-F1 mean of 0.234${\pm}$0.047 while saving 53.2\% of transmitted bytes, compared with 0.209${\pm}$0.013 for the strongest reproduced external baseline.
\end{itemize}

\section{Related Work}

\subsection{Efficient Communication in LLM Multi-Agent Systems}

AutoGen and MetaGPT establish natural-language collaboration among role-specialized agents~\cite{wu2023autogen,hong2023metagpt}. Evolving Idea Graphs retains claims and support/conflict relations in a persistent graph for inspectable scientific ideation~\cite{dong2026eig}, but does not control transmission under resource constraints.

Communication-efficiency methods change the active collaboration structure. Agent, edge, and round pruning suppress participation~\cite{zhang2024agentprune,wang2025agentdropout,safesieve2025,zhang2024sparsedebate}, while architecture search, routing, and subject-level decomposition select task-dependent topologies or agent compositions~\cite{zhang2025gdesigner,hu2025maas,chen2025masrouter,dong2026sdag}. These methods choose who participates or where information flows, not which content crosses a retained route. Prompt compression shortens assembled context~\cite{jiang2023llmlingua,jiang2024longllmlingua,pan2024llmlingua2}; KVCOMM and Q-KVComm reduce model-internal transfer or repeated prefill~\cite{ye2025kvcomm,qkvcomm2025}. BANDMAS instead schedules auditable semantic packets after generation and before receiver admission while leaving topology and models fixed.

\subsection{Semantic and Value-Aware Communication}

Learned multi-agent communication optimizes what, when, and to whom agents transmit~\cite{sukhbaatar2016commnet,foerster2016dial,das2019tarmac,singh2019ic3net,kim2019schednet}, typically learning latent messages jointly with a task policy. Semantic communication prioritizes task value over bit-perfect reconstruction~\cite{xie2021deepsc}; recent systems address QoS-aware gathering, adaptive transmission, lossy distributed inference, token aggregation, and federated LLM training~\cite{wang2024qosscheduling,liao2024adasem,zheng2026halo,cui2025semanticpacket,hu2022distributedinference,su2024titanic}. They optimize channel representations, model updates, or token streams rather than typed fields within independently generated agent messages.

HybridFlow routes dependency-aware subtasks between edge and cloud using learned benefit--cost utility~\cite{dong2025hybridflow}; BANDMAS instead decides which already-generated semantic records enter a receiver using causality-inspired packet utility scores for admission.

BANDMAS obtains causality-inspired supervision through controlled receiver replay motivated by counterfactual removal~\cite{lin2021gem}. Sufficiency and necessity contrasts label packet contribution under fixed conditions; a split-safe predictor then estimates this bounded value from pre-outcome metadata. The online scheduler trades predicted value against measured resource pressure without gold outcomes or a task-specific learned channel. Its causal scope remains limited to the declared isolation and removal contrasts.

\section{Problem Formulation}

We consider a deployed multi-agent workflow whose roles and communication routes are fixed independently of BANDMAS. Producer agents generate natural-language messages from their local observations, and a receiver aggregates the admitted content into a task output. BANDMAS acts only at this receiver-admission boundary: it may suppress or defer generated content, but it does not revise producer messages, invoke an alternative agent, or observe the receiver outcome before making the decision. This separation isolates communication control from topology construction and defines the information available to the scheduler.

For task $x$, a multi-agent workflow emits timestamped messages represented as

\begin{equation}
m_j=(s_j,r_j,y_j,\tau_j,d_j),
\end{equation}

where $s_j$, $r_j$, $y_j$, $\tau_j$, and $d_j$ are the sender, receiver, natural-language payload, emission time, and task-relative deadline. BANDMAS decomposes messages before receiver admission:

\begin{equation}
\mathcal{P}(m_j)=\{p_{j1},\ldots,p_{jk}\}.
\end{equation}

A packet records its message identifier, route, semantic type, timestamp, provenance, segment order, dependency metadata, semantic span, and payload. Controller-side metadata supports validation and scoring, while the compact transmitted record contains only the fields required to route and reconstruct the selected content. Its decision-time resource vector is

\begin{equation}
\mathbf{c}_t(p)=\left(c_{\mathrm{wire}}(p),c_{\mathrm{link},t}(p),c_{\mathrm{ctx}}(p)\right).
\end{equation}

Here, $c_{\mathrm{wire}}$ is the UTF-8 length of the compact application-layer record, $c_{\mathrm{link},t}$ is its predicted transfer delay under current bandwidth and backlog, and $c_{\mathrm{ctx}}$ is its receiver-context token estimate. Wire cost includes the compact packet envelope but excludes transport and link-layer headers. For $\mathcal{P}_x=\cup_j\mathcal{P}(m_j)$, the scheduler chooses a delivered subset $S$ before receiver execution. The valid-set family $\mathcal{V}_x$ enforces routes, deadlines, and the no-outcome-leakage boundary; declared claim--evidence links are retained for audit but not enforced as dependency closure. Full communication sends $\mathcal{P}_x$ under the same serializer and serves as the unconstrained reference.

The three resource components serve different roles. Wire and context costs determine hard per-task feasibility, whereas link delay depends on shared queue state and enters adaptive admission. A packet is deadline-infeasible when its predicted queued transfer and receiver service cannot finish before $d_j$. GPU allocation is measured separately because model residency and KV-cache reuse are serving-stack properties rather than additive packet costs.

Let $F_x(S)$ be the receiver output and let $T(F_x(S),x)$ and $E(F_x(S),x)$ be task and evidence utility. The implemented replay profiler combines the two task-facing metrics as

\begin{equation}
U(F_x(S),x)=\tfrac{1}{2}T(F_x(S),x)+\tfrac{1}{2}E(F_x(S),x).
\end{equation}

SciFact uses label accuracy and rationale F1; HotpotQA and FanOutQA use answer F1 and evidence F1. Equal weighting rewards both correct decisions and retained support. The constrained objective is

\begin{equation}
\begin{aligned}
\max_{S\subseteq\mathcal{P}_x}\quad &U(F_x(S),x)\\
\mathrm{s.t.}\quad
&\sum_{p\in S}c_{\mathrm{wire}}(p)\leq C_{\mathrm{wire}},\\
&\sum_{p\in S}c_{\mathrm{ctx}}(p)\leq C_{\mathrm{ctx}},\quad
S\in\mathcal{V}_x .
\end{aligned}
\end{equation}

BANDMAS labels packet value through bounded replays motivated by counterfactual removal~\cite{lin2021gem}. For $p_i\in\mathcal{P}_x$, define

\begin{equation}
\begin{aligned}
\Delta_{\mathrm{suf}}(p_i)&=U(F_x(\{p_i\}),x)-U(F_x(\emptyset),x),\\
\Delta_{\mathrm{nec}}(p_i)&=U(F_x(\mathcal{P}_x),x)-U(F_x(\mathcal{P}_x\!\setminus\!\{p_i\}),x),\\
\Delta(p_i)&=\tfrac{1}{2}\bigl(\Delta_{\mathrm{suf}}(p_i)+\Delta_{\mathrm{nec}}(p_i)\bigr).
\end{aligned}
\end{equation}

We call $\Delta(p_i)$ the replay causal effect (RCE). Every contrast fixes the task, producer messages, receiver, prompt, decoding policy, and utility. Sufficiency rewards independently useful content; necessity captures unique contribution in the complete context. Positive, near-zero, and negative values denote helpful, redundant, and harmful packets under these declared contexts. RCE is a context-dependent, causality-inspired profiling label, not an unrestricted causal effect or online outcome signal, and it omits arbitrary higher-order interactions.

The two contrasts address complementary failure modes. Necessity can be zero when several packets are substitutable, even if each is useful alone; sufficiency can undervalue evidence that becomes useful only in combination. Their average is therefore a bounded ranking target rather than an additive decomposition of receiver utility. Packet interactions beyond the two declared contexts remain outside the estimator's target.

At deployment, $\Delta(p_i)$ is unavailable because it requires replayed receiver outcomes. BANDMAS therefore freezes an estimator learned from pre-outcome metadata before final-test scheduling. Decisions may use generated packets, the frozen estimator, and completed resource observations, but not gold labels, receiver answers, evaluation scores, or final-test RCE.

\section{BANDMAS Design}
\label{sec:bandmas-design}

\begin{figure}[t]
\centering
\includegraphics[width=\columnwidth]{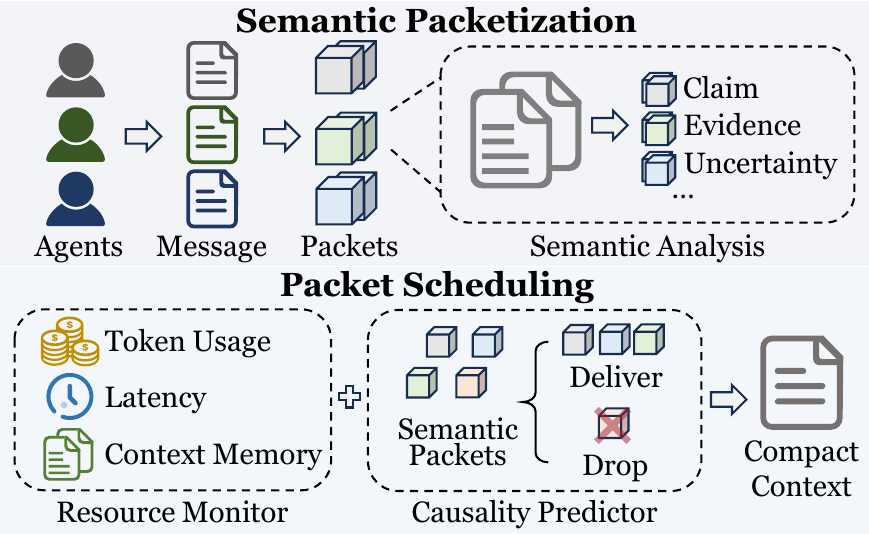}
\caption{BANDMAS online framework. Generated messages become typed semantic packets; a replay-trained predictor and resource monitor select complete records for the receiver context.}
\label{fig:bandmas-framework}
\end{figure}

BANDMAS interposes between message generation and receiver inference without changing agent roles or routes. Figure~\ref{fig:bandmas-framework} shows its online path: packetize generated messages, score packets with a frozen replay-trained predictor, and admit complete records under current resource constraints. Replay profiling occurs offline.

\subsection{End-to-End Control Path}

Each producer emits a source-grounded structured message under a common prompt contract. The packetizer constructs typed records and computes serialized wire cost; the frozen scorer then assigns a predicted contribution using only pre-outcome metadata. The scheduler admits records until a wire, context, or deadline constraint becomes active, after which the receiver runs once on the selected content. Offline profiling replays disjoint profile messages, computes RCE labels, fits the shrinkage predictor, and freezes it before final-test scheduling.

This separation keeps expensive counterfactual receiver calls off the deployment path. Online scoring is a metadata lookup through the frozen hierarchy, and scheduling uses one ranking pass plus exact serialized-batch accounting. Profile and final-test partitions share the packet schema and receiver implementation, but never replay outcomes.

\subsection{Semantic Packet Interface}

Reader agents emit one source-grounded JSON object with Claim, Evidence, Uncertainty, and Request strings. A deterministic parser converts these fields into a canonical message; parse failures produce the same four fields from source-grounded fallback content and are recorded. The packetizer splits at typed field markers, falling back to sentences when markers are absent and merging fragments shorter than three words into the preceding packet. Each packet records its message and segment identifiers, route, timestamp, deadline, type, provenance, packetization method, and semantic span. Claim packets retain same-message evidence identifiers for audit, although scheduling does not enforce dependency closure. The complete ordered packet set reconstructs the canonical message, making Full communication a lossless reference.

The four fields serve distinct communication roles. Claim presents the conclusion backed by sources, Evidence identifies or quotes its supporting material, Uncertainty records confidence limitations or unresolved ambiguity, and Request identifies information needed from another participant or the receiver. Field-level word limits prevent the structured contract from becoming a second unrestricted message, and all content must remain grounded in the reader's assigned source.

Packets are charged after packetization using a compact UTF-8 JSONL serializer. A transmitted batch contains task, route, message and grouping identifiers, segment order, semantic type, and content; richer controller metadata remains in the experiment trace. Wire cost therefore includes the packet envelope rather than payload characters alone, exposing the overhead created by finer packetization.

Admission always evaluates the exact serialized batch rather than summing payload lengths. This captures shared batch fields, escaping, and per-record envelope overhead. The receiver reconstructs selected records in message and segment order, so packet scheduling changes admitted content but not the representation seen by different policies.

\subsection{Causality-Inspired, Split-Safe Replay Valuation}

Profiling evaluates the Full, Empty, Full-without-packet, and packet-alone contexts used in the RCE definition. Full and Empty are reused within each task. Profiling is restricted to profile-train and validation partitions; malformed traffic or a benchmark with no positive replay effect fails promotion. The predictor is then frozen before final-test packetization. Final-test outcomes, labels, and RCE values are forbidden features.

The split-safe feature set contains packet type, route, lexical markers, serialized size, and turn position. It excludes receiver outputs and any feature computed from task or evidence scores. This design trades predictive flexibility for an auditable deployment boundary and avoids an online LLM judge whose latency and nondeterminism would confound the communication measurements.

The predictor uses empirical-Bayes-style shrinkage with strength $\alpha=6$. For a feature cell $g$ with support $n_g$, empirical mean $\bar{\Delta}_g$, and broader prior $\pi_g$, the recursive estimate is

\begin{equation}
\widehat{\mu}_g=
\frac{n_g}{n_g+\alpha}\bar{\Delta}_g+
\frac{\alpha}{n_g+\alpha}\pi_g.
\label{eq:shrinkage-predictor}
\end{equation}

The hierarchy begins at the global RCE mean, forms packet-type and route priors, and refines them with type--route, lexical-marker, packet-size, and turn cells. Parallel type--route and marker priors are averaged before the most specific cell. The deployed estimate $\widehat{\Delta}(p)=\widehat{\mu}_{g(p)}$ remains traceable to observable metadata and support counts. It predicts bounded replay contribution, not a final-test outcome or an effect beyond the replay design.

Shrinkage is important because fine-grained cells may receive little profile support. Equation~\eqref{eq:shrinkage-predictor} shifts sparse estimates toward broader priors while allowing well-supported cells to retain their empirical means. The fixed strength $\alpha=6$ is shared across benchmarks.

BANDMAS exposes two scoring policies:

\begin{equation}
\begin{aligned}
s_{\mathrm{causal}}(p) &= \widehat{\Delta}(p),\\
s_{\mathrm{hybrid}}(p) &=
\lambda_1\operatorname{Norm}_{x}\!\left(\widehat{\Delta}(p)\right)
+\lambda_2 q_{\mathrm{sem}}(p),
\end{aligned}
\quad \lambda_1+\lambda_2=1,
\label{eq:bandmas-policy-scores}
\end{equation}

where $\operatorname{Norm}_{x}(\cdot)$ is task-level min--max normalization and $q_{\mathrm{sem}}(p)$ is a pre-outcome semantic prior. The reported hybrid configuration sets $\lambda_1=0.75$ and $\lambda_2=0.25$; when no separate relevance score is stored for structured generated traffic, $q_{\mathrm{sem}}$ is the evidence-type indicator. Measured RCE never appears in a final-test decision.

\subsection{Static and Adaptive Packet Scheduling}

Static BANDMAS sorts packets by frozen score and greedily admits each complete record when the resulting serialized batch remains within the task's wire cap. Packet identifiers break ties, and skipped packets do not consume capacity. Recomputing batch size after each candidate captures shared-envelope effects. Full communication bypasses the cap under the same serializer.

Adaptive BANDMAS retains the predictor and hard wire cap but adds deadline, backlog, context, and price checks. Resource prices follow

\begin{equation}
\lambda_{k,t+1}=\max\!\left(0,\lambda_{k,t}+\eta_k\left(\frac{u_{k,t}}{C_{k,t}}-1\right)\right),
\end{equation}

where $u_{k,t}$ is completed wire volume, service latency, or receiver-context use. Before each arrival, the controller processes only completed transfer and receiver observations. It admits packets with positive net value after byte, predicted-delay, and context charges. It discards expired, cap-violating, or deadline-infeasible packets. Packets blocked by context constraints are deferred.

Prices rise when observed use exceeds the corresponding capacity and decay otherwise. Because updates use completed service only, the decision for one arrival cannot depend on future queue outcomes. Static and adaptive modes share packet values and the hard wire cap; adaptation changes only the resource-sensitive admission threshold.

\begin{algorithm}[t]
\caption{Online BANDMAS packet admission}
\label{alg:bandmas-admission}
\footnotesize
\begin{algorithmic}[1]
\REQUIRE Messages $\mathcal{M}_x$, frozen predictor $f$, policy $h$, state $z_t$
\ENSURE Admitted packet set $S$
\STATE $\mathcal{P}_x \leftarrow \textsc{SemanticPacketize}(\mathcal{M}_x)$
\STATE $s(p) \leftarrow s_h(p;f)$ for each $p\in\mathcal{P}_x$
\IF{adaptive mode}
    \STATE update $\lambda_{k,t}$ from completed service observations
\ENDIF
\STATE $S\leftarrow\emptyset$; rank $\mathcal{P}_x$ by $s(p)$, then packet ID
\FOR{each packet $p$ in ranked order}
    \STATE compute exact serialized bytes of $S\cup\{p\}$
    \IF{the serialized batch exceeds the wire cap}
        \STATE drop $p$
    \ELSIF{adaptive mode and $p$ is expired or deadline-infeasible}
        \STATE drop $p$
    \ELSIF{adaptive mode and $p$ exceeds the context limit}
        \STATE defer $p$
    \ELSIF{static mode or $g_t(p)>0$}
        \STATE $S\leftarrow S\cup\{p\}$
    \ELSE
        \STATE drop $p$
    \ENDIF
\ENDFOR
\RETURN $S$ in message and segment order
\end{algorithmic}
\end{algorithm}

Algorithm~\ref{alg:bandmas-admission} unifies the static and adaptive paths. For adaptive admission, the net value is

\begin{equation}
g_t(p)=s_h(p)-
\lambda_{b,t}\frac{c_{\mathrm{wire}}(p)}{C_b}
-\lambda_{\ell,t}\frac{c_{\mathrm{link},t}(p)}{C_\ell}
-\lambda_{q,t}\frac{c_{\mathrm{ctx}}(p)}{C_q}.
\label{eq:adaptive-net-value}
\end{equation}

Ranking costs $O(n\log n)$ for $n$ candidates. Exact serialized-batch accounting after each candidate yields an $O(n^2)$ worst-case admission pass; Section~V measures this overhead separately from inference. Offline replay supplies task-grounded labels, whereas final-test scheduling uses only frozen predictions and pre-admission resource observations.

The current implementation consumes routes exposed by the active workflow instead of generating a collaboration topology. This boundary makes the controller compatible with existing workflows, but topology adaptation and enforced dependency closure require separate mechanisms.

\section{Evaluation}

We evaluate end-task utility under constrained communication conditions, replay-derived packet scores, packetization, and system behavior under controlled load levels. All policies use fixed producer traffic, which separates communication control processes from message generation.

\subsection{Experimental Setup}

\textbf{Datasets and task design.} SciFact tests source-grounded claim verification over scientific abstracts~\cite{wadden2020scifact}. HotpotQA requires complementary bridge evidence from multiple documents~\cite{yang2018hotpotqa}, while FanOutQA distributes useful evidence across a wider reader set~\cite{zhu2024fanoutqa}. The suite thus spans verification, multi-hop composition, and high-fan-out aggregation. Each split seed contains 160 final-test tasks for SciFact, 160 for HotpotQA, and 98 for FanOutQA.

The benchmarks are selected for their distinct communication structures. SciFact may be solved using one decisive rationale. HotpotQA requires combining information from multiple separate sources. FanOutQA focuses on picking information from a larger reader set. This variation tests whether packet value remains useful as evidence becomes more distributed.

\begin{table*}[t]
    \centering
    \caption{Main results (mean$\pm$sample standard deviation, three split seeds). Tx bytes are compact application-layer records after packetization and admission; latency is warmed receiver time.}
    \label{tab:main-results}
    \small
    \setlength{\tabcolsep}{1.5pt}
    \begin{tabular*}{\textwidth}{@{\extracolsep{\fill}}llcrrrrr@{}}
        \toprule
        & & & \multicolumn{2}{c}{Utility} & \multicolumn{3}{c}{System} \\
        \cmidrule(lr){4-5}\cmidrule(lr){6-8}
        Benchmark & Baseline & Budget & Task & Evidence & Tx bytes & Saving (\%) & \shortstack{Latency\\(ms)} \\
        \midrule
        SciFact & Full communication & 1.00 & 0.617{\scriptsize ${\pm}$0.030} & 0.397{\scriptsize ${\pm}$0.023} & 1099.2{\scriptsize ${\pm}$49.8} & -- & 620.3{\scriptsize ${\pm}$2.0} \\
SciFact & AgentPrune & 0.50 & 0.640{\scriptsize ${\pm}$0.010} & 0.373{\scriptsize ${\pm}$0.020} & 413.3{\scriptsize ${\pm}$28.7} & 62.4 & 623.0{\scriptsize ${\pm}$5.0} \\
SciFact & SafeSieve & 0.50 & 0.635{\scriptsize ${\pm}$0.022} & 0.362{\scriptsize ${\pm}$0.011} & 422.7{\scriptsize ${\pm}$19.4} & 61.5 & 625.5{\scriptsize ${\pm}$3.2} \\
SciFact & Semantic relevance & 0.50 & 0.425{\scriptsize ${\pm}$0.027} & 0.386{\scriptsize ${\pm}$0.040} & 483.9{\scriptsize ${\pm}$26.9} & 57.0 & 620.1{\scriptsize ${\pm}$3.2} \\
SciFact & BANDMAS hybrid & 0.50 & 0.665{\scriptsize ${\pm}$0.007} & 0.375{\scriptsize ${\pm}$0.019} & 410.4{\scriptsize ${\pm}$24.8} & 63.6 & 616.1{\scriptsize ${\pm}$3.5} \\
\bandmasrowfill{\textwidth}SciFact & BANDMAS causal & 0.50 & \textbf{0.679{\scriptsize ${\pm}$0.013}} & \textbf{0.398{\scriptsize ${\pm}$0.021}} & \textbf{408.8{\scriptsize ${\pm}$25.5}} & \textbf{64.2} & \textbf{615.3{\scriptsize ${\pm}$5.3}} \\
\midrule
HotpotQA & Full communication & 1.00 & 0.273{\scriptsize ${\pm}$0.038} & 0.207{\scriptsize ${\pm}$0.011} & 4394.3{\scriptsize ${\pm}$18.7} & -- & 702.9{\scriptsize ${\pm}$6.5} \\
HotpotQA & AgentPrune & 0.25 & 0.245{\scriptsize ${\pm}$0.047} & 0.164{\scriptsize ${\pm}$0.023} & 1004.9{\scriptsize ${\pm}$7.3} & 77.0 & 637.7{\scriptsize ${\pm}$5.2} \\
HotpotQA & SafeSieve & 0.25 & 0.250{\scriptsize ${\pm}$0.014} & 0.199{\scriptsize ${\pm}$0.017} & 1006.4{\scriptsize ${\pm}$1.0} & 77.1 & 641.7{\scriptsize ${\pm}$4.2} \\
HotpotQA & Semantic relevance & 0.25 & 0.134{\scriptsize ${\pm}$0.009} & 0.050{\scriptsize ${\pm}$0.010} & 1016.5{\scriptsize ${\pm}$11.4} & 76.9 & 676.6{\scriptsize ${\pm}$5.1} \\
HotpotQA & BANDMAS hybrid & 0.25 & 0.289{\scriptsize ${\pm}$0.021} & 0.188{\scriptsize ${\pm}$0.016} & 998.1{\scriptsize ${\pm}$1.6} & 77.2 & 628.0{\scriptsize ${\pm}$9.7} \\
\bandmasrowfill{\textwidth}HotpotQA & BANDMAS causal & 0.25 & \textbf{0.290{\scriptsize ${\pm}$0.021}} & \textbf{0.212{\scriptsize ${\pm}$0.016}} & \textbf{997.9{\scriptsize ${\pm}$1.7}} & \textbf{77.3} & \textbf{626.4{\scriptsize ${\pm}$8.6}} \\
\midrule
FanOutQA & Full communication & 1.00 & 0.219{\scriptsize ${\pm}$0.009} & 0.120{\scriptsize ${\pm}$0.006} & 4320.5{\scriptsize ${\pm}$20.5} & -- & 782.6{\scriptsize ${\pm}$3.5} \\
FanOutQA & AgentPrune & 0.50 & 0.113{\scriptsize ${\pm}$0.005} & 0.123{\scriptsize ${\pm}$0.011} & 2038.6{\scriptsize ${\pm}$8.4} & 52.8 & 744.4{\scriptsize ${\pm}$4.6} \\
FanOutQA & SafeSieve & 0.50 & 0.209{\scriptsize ${\pm}$0.013} & 0.174{\scriptsize ${\pm}$0.006} & 2043.6{\scriptsize ${\pm}$11.3} & 52.7 & 748.8{\scriptsize ${\pm}$6.8} \\
FanOutQA & Semantic relevance & 0.50 & 0.116{\scriptsize ${\pm}$0.009} & 0.147{\scriptsize ${\pm}$0.014} & 2058.8{\scriptsize ${\pm}$15.5} & 52.4 & 744.1{\scriptsize ${\pm}$4.9} \\
FanOutQA & BANDMAS hybrid & 0.50 & 0.228{\scriptsize ${\pm}$0.036} & 0.240{\scriptsize ${\pm}$0.012} & 2032.3{\scriptsize ${\pm}$6.9} & 53.0 & \textbf{738.7{\scriptsize ${\pm}$6.3}} \\
\bandmasrowfill{\textwidth}FanOutQA & BANDMAS causal & 0.50 & \textbf{0.234{\scriptsize ${\pm}$0.047}} & \textbf{0.257{\scriptsize ${\pm}$0.011}} & \textbf{2028.3{\scriptsize ${\pm}$6.1}} & \textbf{53.2} & 740.8{\scriptsize ${\pm}$5.3} \\
 
        \bottomrule
    \end{tabular*}
\end{table*}

\textbf{Metrics.} Task utility is label accuracy for SciFact and answer F1 for HotpotQA and FanOutQA. Evidence utility is rationale, supporting-fact, or evidence F1, preventing traffic reduction from appearing beneficial when support is lost. Communication cost is the UTF-8 size of compact post-admission records, including packet envelopes but excluding transport headers; savings use same-task Full communication. We also report warmed receiver latency, receiver-context tokens, scheduler overhead, p95 latency, and deadline misses to distinguish network savings from shifted inference cost.

Task and evidence metrics are reported separately even though their equal-weighted combination supervises replay profiling. This prevents the training objective from hiding whether a policy preserves final-answer quality, supporting evidence, or both. Latency is measured after receiver warm-up and is not interpreted as network delay in the main table.

\textbf{Baselines.} Full sends the complete packet stream as an unconstrained reference. Semantic relevance ranks packets by lexical alignment without replay supervision. AgentPrune~\cite{zhang2024agentprune} represents communication-graph pruning, and SafeSieve~\cite{safesieve2025} represents semantic and experience-guided progressive pruning. Both are official-code-derived reproductions adapted to the frozen packet stream, receiver, evaluator, and wire cap, rather than native end-to-end repository executions. \emph{BANDMAS causal} uses predicted RCE, not online causal estimation; hybrid follows~\eqref{eq:bandmas-policy-scores}. Prompt compressors are excluded because they operate after receiver-context assembly.

For fairness, every constrained method is assigned the identical task-specific cap and pre-generated traffic. The external reproductions retain their native selection logic while making this logic accessible via the common packet, serialization, and evaluation protocol. Thus, differences in Table~\ref{tab:main-results} arise from admission decisions rather than alternative prompts, model outputs, or byte accounting.

\textbf{Implementation details.} Each reader receives one source and emits the four structured fields under a common grounding contract. We serve Q4\_K\_M \texttt{qwen3:4b} through Ollama~0.32.1 on a 24-GB RTX~4090 with temperature zero, disabled thinking, strict JSON, a 256-token cap, and no response cache. Temperature zero freezes producer traffic rather than measuring decoding variability. Prompts, messages, fallbacks, provenance, and corpus hashes are fixed before packetization; gold answers and evidence are hidden from producers. Three seeds vary task selection and profile/validation/final-test membership. Replay labels use only profile and validation tasks, and the predictor is frozen before final-test scheduling. Latency and communication are recorded per task; GPU telemetry is sampled every 100~ms.

The receiver uses the same prompt contract and decoding settings across policies. Parser fallbacks are frozen with the message corpus, ensuring that packetization does not regenerate or repair a policy-specific message. Seeds vary data partitions, not stochastic decoding, so reported variability reflects task composition under identical generated traffic.

\textbf{Budgets and statistics.} Each constrained policy receives $C_{\mathrm{wire}}=\lfloor\beta B_{\mathrm{full}}\rfloor$, where $\beta\in\{0.25,0.50,0.75\}$ and $B_{\mathrm{full}}$ is same-task Full traffic. Table~\ref{tab:main-results} uses predeclared caps 0.50, 0.25, and 0.50 for SciFact, HotpotQA, and FanOutQA; frontiers show the sweep. Hybrid sets $(\lambda_1,\lambda_2)=(0.75,0.25)$. We report mean and sample standard deviation across three split seeds. Comparisons pair tasks, seeds, corpora, and caps. Repeated tasks are averaged before task-cluster bootstrap intervals and paired sign-flip tests, with Holm correction across six causal-versus-semantic task/evidence tests. The task, not the packet, is the inferential unit.

Task-cluster resampling preserves within-task dependence among packets and repeated evaluations. Paired sign flips test the mean task-level policy difference under the shared traffic and cap; Holm correction controls family-wise error across the declared task/evidence comparisons. Cross-policy means without a corresponding paired test are described as rankings rather than significance claims.

\subsection{Main Results}

Table~\ref{tab:main-results} compares policies at each benchmark's predeclared cap. Full establishes the unconstrained utility and traffic scale; all methods use the same generated messages and receiver.

\textbf{Causal admission has the highest task-utility mean at every reported cap.} Its means are 0.679 on SciFact, 0.290 on HotpotQA, and 0.234 on FanOutQA, compared with Full means of 0.617, 0.273, and 0.219 while saving 64.2\%, 77.3\%, and 53.2\% of bytes. Hybrid is close on HotpotQA (0.289), and the FanOutQA margin over Full is small. Rankings against baselines other than semantic relevance remain descriptive because the paired tests below do not cover every policy pair.

Against reproduced external baselines, causal exceeds AgentPrune and SafeSieve on all three task metrics. The largest performance gap with external baselines appears on HotpotQA (0.290 versus 0.250 for SafeSieve), while FanOutQA is narrower (0.234 versus 0.209 for SafeSieve). These means show that the byte savings do not arise from indiscriminate suppression, but three split seeds are insufficient for claims about every cross-policy difference.

\textbf{Evidence behavior depends on benchmark structure.} SciFact evidence-F1 means are nearly tied (0.398 causal versus 0.397 Full), and causal's paired difference from semantic relevance is non-significant. Causal improves evidence F1 over semantic relevance by 0.144 on HotpotQA and 0.034 on FanOutQA; both intervals exclude zero after Holm correction. Replay-derived value is most useful here when complementary evidence is distributed across sources.

This distinction matters because task F1 can remain stable when the receiver reaches the right answer from incomplete support. The evidence column therefore bounds the interpretation of the task gains: BANDMAS preserves or improves support on the multi-source benchmarks, but does not improve SciFact rationale quality over Full in a meaningful way.

\begin{figure*}[t]
\centering
\includegraphics[width=0.96\textwidth]{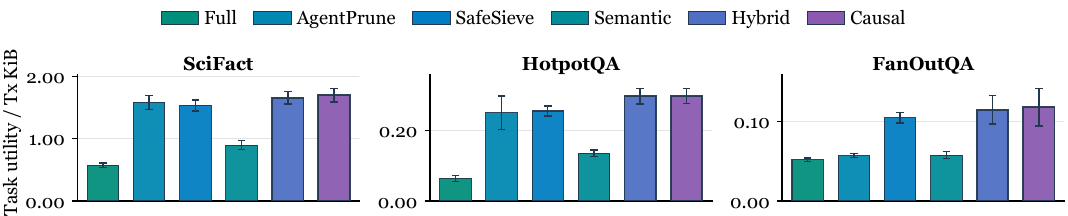}
\caption{Task utility per transmitted application-layer KiB ($1$~KiB $=1024$ bytes) at the Table~\ref{tab:main-results} operating points. Bars use the table means; error bars apply first-order propagation to the reported task and byte standard deviations with zero covariance. Absolute utility and traffic remain in Table~\ref{tab:main-results}.}
\label{fig:system-efficiency}
\end{figure*}

\textbf{Utility per transmitted KiB summarizes the trade-off.} Figure~\ref{fig:system-efficiency} divides task utility by post-admission traffic. Causal reaches 1.701, 0.297, and 0.118 utility/KiB on SciFact, HotpotQA, and FanOutQA, respectively. It outperforms the strongest reproduced baseline across every benchmark. This ratio complements the absolute metrics because packet envelopes and indivisible records create differences in realized traffic below a common cap.

The ratio serves as an operating-point summary, not a substitute for the utility--cost frontier: it may heavily favor an extremely small transmission even when absolute utility is poor. We therefore retain Table~\ref{tab:main-results} and the packetization frontier as our core supporting evidence.

\textbf{Byte reduction does not imply proportional inference speedup.} Constrained policies save 52.4--77.3\% of bytes, but warmed receiver latency falls less and depends on policy. Causal is fastest on SciFact and HotpotQA, while hybrid is fastest on FanOutQA; these measurements separate network efficiency from receiver computation rather than establish universal latency gains.

\subsection{Causality-Inspired Replay Scoring Analysis}

Table~\ref{tab:paired-deltas} compares causal admission with semantic relevance using task-level paired inference at the same cap. Table~\ref{tab:causal-validation} reports replay-label composition, rank agreement, and useful-packet recall on profile/validation traffic.

\begin{table}[t]
\centering
\caption{Paired causal-minus-semantic effects at representative caps. CIs use task-cluster bootstrap; $p_{\rm H}$ uses Holm-adjusted sign-flip tests.}
\label{tab:paired-deltas}
\small
\setlength{\tabcolsep}{1.0pt}
\begin{tabular*}{\columnwidth}{@{\extracolsep{\fill}}llrrr@{}}
\toprule
Benchmark & Metric & $\Delta$ [95\% CI] & $p_{\rm H}$ & $\Delta$ bytes \\
\midrule
SciFact & Task & ${+}$0.236 [${+}$0.171, ${+}$0.299] & $<$0.001 & ${-}$67.7 \\
 & Evidence & ${-}$0.018 [${-}$0.064, ${+}$0.027] & 0.418 & -- \\
\addlinespace[1pt]
HotpotQA & Task & ${+}$0.157 [${+}$0.120, ${+}$0.195] & $<$0.001 & ${-}$17.9 \\
 & Evidence & ${+}$0.144 [${+}$0.116, ${+}$0.171] & $<$0.001 & -- \\
\addlinespace[1pt]
FanOutQA & Task & ${+}$0.105 [${+}$0.081, ${+}$0.132] & $<$0.001 & ${-}$26.9 \\
 & Evidence & ${+}$0.034 [${+}$0.007, ${+}$0.061] & 0.026 & -- \\
 
\bottomrule
\end{tabular*}
\end{table}

\begin{table}[t]
\centering
\caption{Replay-label composition and frozen-scorer diagnostics on profile/validation traffic.}
\label{tab:causal-validation}
\small
\setlength{\tabcolsep}{0.8pt}
\begin{tabular*}{\columnwidth}{@{\extracolsep{\fill}}lrrrrr@{}}
\toprule
Benchmark & Pos. (\%) & Zero (\%) & Neg. (\%) & \shortstack{Spearman\\$\rho$} & \shortstack{Useful\\recall} \\
\midrule
SciFact & 34.8 & 51.4 & 13.7 & 0.481{\scriptsize ${\pm}$0.033} & 0.522{\scriptsize ${\pm}$0.031} \\
HotpotQA & 36.0 & 58.3 & 5.7 & 0.538{\scriptsize ${\pm}$0.015} & 0.582{\scriptsize ${\pm}$0.025} \\
FanOutQA & 35.1 & 59.4 & 5.5 & 0.480{\scriptsize ${\pm}$0.013} & 0.547{\scriptsize ${\pm}$0.032} \\
 
\bottomrule
\end{tabular*}
\end{table}

\textbf{Replay valuation improves task utility over semantic relevance at matched caps.} Paired performance gains are 0.236 on SciFact, 0.157 on HotpotQA, and 0.105 on FanOutQA; all intervals exclude zero and Holm-adjusted task tests have $p<0.001$. The Causal setup also transmits 17.9--67.7 fewer bytes. Evidence gains are significant on HotpotQA and FanOutQA, whereas SciFact changes by $-0.018$ with an interval spanning zero. The tests support predicted RCE over semantic ranking within the replay design, not unrestricted causal effects or superiority over every baseline.

\textbf{The frozen predictor is informative but imperfect.} Positive replay labels account for 34.8--36.0\% of packets and 51.4--59.4\% have zero measured contribution. Cross-fitted rank correlation is 0.480--0.538 and useful-packet recall is 0.522--0.582, supporting RCE as a scheduling signal but not an exact, context-independent ordering.

The large zero class explains why relevance alone is insufficient: many topically aligned units do not change measured receiver utility in either declared replay context. Moderate rank correlation also explains why hybrid scoring remains competitive on some benchmarks; the predicted RCE ordering contains useful but incomplete information.

\subsection{Model Transfer}

Table~\ref{tab:qwen25-transfer} repeats the constrained comparison with Qwen2.5-7B-Instruct on SciFact and HotpotQA. The runs use frozen matched traffic and three split seeds, but different caps and serving software from the Qwen3-4B study, so they are a transfer check rather than pooled main results.

\begin{table*}[t]
\centering
\caption{Qwen2.5-7B transfer (mean$\pm$sample standard deviation, three split seeds). Caps are 0.50 for SciFact and 0.35 for HotpotQA.}
\label{tab:qwen25-transfer}
\small
\setlength{\tabcolsep}{1.0pt}
\begin{tabular*}{\textwidth}{@{\extracolsep{\fill}}lrrrlrrr@{}}
\toprule
\multicolumn{4}{c}{SciFact ($\beta=0.50$)} & \multicolumn{4}{c}{HotpotQA ($\beta=0.35$)} \\
\cmidrule(lr){1-4}\cmidrule(lr){5-8}
Baseline & Task & Evidence & Tx bytes & Baseline & Task & Evidence & Tx bytes \\
\midrule
AgentPrune & 0.412{\scriptsize ${\pm}$0.017} & 0.177{\scriptsize ${\pm}$0.022} & 2065.0{\scriptsize ${\pm}$80.0} & AgentPrune & 0.458{\scriptsize ${\pm}$0.027} & 0.446{\scriptsize ${\pm}$0.033} & \textbf{3172.5{\scriptsize ${\pm}$43.0}} \\
SafeSieve & 0.694{\scriptsize ${\pm}$0.016} & 0.526{\scriptsize ${\pm}$0.012} & 2069.3{\scriptsize ${\pm}$77.8} & SafeSieve & 0.539{\scriptsize ${\pm}$0.027} & 0.513{\scriptsize ${\pm}$0.014} & 3310.4{\scriptsize ${\pm}$70.1} \\
Semantic rel. & 0.683{\scriptsize ${\pm}$0.026} & 0.443{\scriptsize ${\pm}$0.026} & 1997.5{\scriptsize ${\pm}$82.7} & Semantic rel. & 0.547{\scriptsize ${\pm}$0.048} & 0.525{\scriptsize ${\pm}$0.017} & 3213.7{\scriptsize ${\pm}$99.4} \\
\bandmasrowfill{\textwidth}Causal & \textbf{0.702{\scriptsize ${\pm}$0.010}} & \textbf{0.533{\scriptsize ${\pm}$0.016}} & \textbf{1857.3{\scriptsize ${\pm}$82.6}} & Causal & \textbf{0.567{\scriptsize ${\pm}$0.048}} & \textbf{0.535{\scriptsize ${\pm}$0.017}} & 3180.7{\scriptsize ${\pm}$99.4} \\
 
\bottomrule
\end{tabular*}
\end{table*}

\textbf{The causal ordering persists under the second model.} Causal has the highest task and evidence means on both benchmarks. It also uses the fewest bytes on SciFact; on HotpotQA it uses 8.2 more bytes than AgentPrune while improving task F1 from 0.458 to 0.567. The check does not cover FanOutQA or paired significance tests.

\subsection{Semantic Packetization Ablation}

We compare three packet boundaries over the same frozen reader outputs: indivisible messages, deterministic sentences, and typed Claim, Evidence, Uncertainty, and Request fields. The lexical selector, receiver, and task-specific wire caps are fixed, excluding replay-value scoring and generation differences while retaining the packet-size and envelope consequences of each boundary.

\begin{figure*}[t]
\centering
\includegraphics[width=0.96\textwidth]{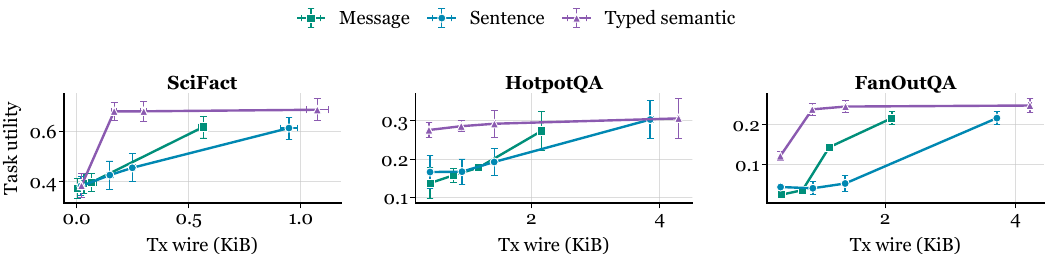}
\caption{Packetization utility--wire frontiers under identical messages and lexical selection. Points denote budgets 0.25, 0.50, 0.75, and Full; bars show one sample standard deviation. Wire traffic includes compact packet envelopes.}
\label{fig:packetization-frontier}
\end{figure*}

\textbf{Typed boundaries have the highest utility means throughout the tested frontier.} At the main caps, typed packets reach 0.679 on SciFact, 0.275 on HotpotQA, and 0.238 on FanOutQA, versus sentence values of 0.425, 0.167, and 0.040 and message values of 0.392, 0.138, and 0.035. Their finer envelopes do not uniformly minimize bytes.

Because generation and lexical selection are fixed, the ablation measures the packetization interface as a whole, including granularity, envelope overhead, and which indivisible units fit. It does not isolate labels from packet-size admissibility.

Message packets are often too large to fit a residual cap, while sentence packets can split a claim from its support. Typed fields provide an intermediate granularity aligned with the reader contract. Their advantage is therefore both semantic and combinatorial: the scheduler receives meaningful units that can be packed without reverting to arbitrary token truncation.

\subsection{Adaptive-Load Operating Envelope and Overhead}

We replay paired Poisson arrivals over $\rho\in\{0.50,0.80,1.00,1.20,1.50\}$, normalized by the slower Full-reference link or receiver capacity, using a 25-kb/s link, one FCFS receiver slot, and a 2.5-s deadline. Twenty disjoint profile/validation tasks per benchmark and seed calibrate capacity and warm queues but are excluded from metrics. Full, semantic, and causal-static keep packet decisions fixed; causal-adaptive uses only resource observations completed before each arrival. This isolates queue-aware control rather than concurrent multi-node deployment.

\begin{figure*}[t]
\centering
\includegraphics[width=0.96\textwidth]{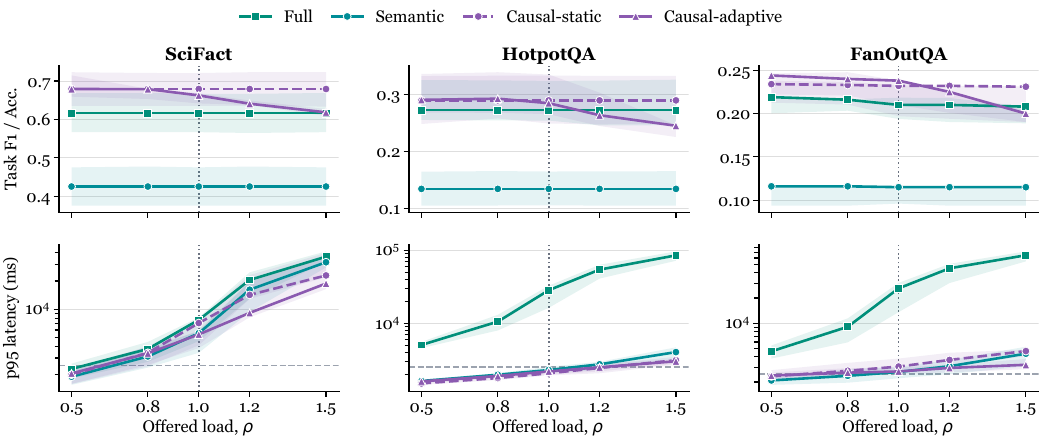}
\caption{Task utility and p95 end-to-end latency over normalized offered load. Curves aggregate three split seeds; bands are 95\% hierarchical-bootstrap intervals. The dotted vertical line marks $\rho=1$, and the dashed horizontal line marks the 2.5-s deadline.}
\label{fig:adaptive-load}
\end{figure*}

\begin{table}[t]
\centering
\caption{System behavior at main operating points. No measured task exceeds the 2.5-s deadline.}
\label{tab:system-overhead}
\small
\setlength{\tabcolsep}{0.5pt}
\begin{tabular*}{\columnwidth}{@{\extracolsep{\fill}}llrrr@{}}
\toprule
Benchmark & Baseline & \shortstack{Context\\tok.} & \shortstack{Sched.\\ms} & \shortstack{p95\\ms} \\
\midrule
SciFact & Full & 53.7{\scriptsize ${\pm}$1.6} & \textbf{1.26{\scriptsize ${\pm}$0.02}} & 745.7{\scriptsize ${\pm}$15.5} \\
\bandmasrowfill{\columnwidth}SciFact & Causal & \textbf{23.5{\scriptsize ${\pm}$0.8}} & 1.93{\scriptsize ${\pm}$0.10} & \textbf{701.7{\scriptsize ${\pm}$1.9}} \\
\midrule
HotpotQA & Full & 232.1{\scriptsize ${\pm}$2.1} & \textbf{1.87{\scriptsize ${\pm}$0.06}} & 985.0{\scriptsize ${\pm}$16.5} \\
\bandmasrowfill{\columnwidth}HotpotQA & Causal & \textbf{68.0{\scriptsize ${\pm}$0.1}} & 4.65{\scriptsize ${\pm}$0.30} & \textbf{730.1{\scriptsize ${\pm}$14.8}} \\
\midrule
FanOutQA & Full & 229.3{\scriptsize ${\pm}$1.1} & \textbf{1.92{\scriptsize ${\pm}$0.09}} & 1439.3{\scriptsize ${\pm}$138.7} \\
\bandmasrowfill{\columnwidth}FanOutQA & Causal & \textbf{122.5{\scriptsize ${\pm}$1.4}} & 6.63{\scriptsize ${\pm}$0.21} & \textbf{1331.3{\scriptsize ${\pm}$26.2}} \\
 
\bottomrule
\end{tabular*}
\end{table}

\textbf{Adaptation matters after queues build.} At $\rho=0.50$, adaptive and static causal have the same SciFact task mean, while adaptive p95 is slightly higher. At $\rho=1.00$, adaptation lowers SciFact p95 from 7,102 to 5,366~ms as accuracy changes from 0.679 to 0.663; FanOutQA p95 falls from 3,054 to 2,672~ms while task F1 rises from 0.232 to 0.238. At $\rho=1.50$, p95 falls from 22,923 to 18,793~ms on SciFact and from 4,668 to 3,207~ms on FanOutQA, but task means fall to 0.618 and 0.200. HotpotQA changes less. Full develops the steepest latency growth because its larger records saturate the link. These operating-envelope comparisons are descriptive because task-level rerun pairs are unavailable.

The operating envelope divides two operational regimes. Below saturation, static ranking already satisfies most deadlines and adaptation provides little benefit. Above saturation, completed-service prices suppress more packets and reduce queue growth, but the selected set becomes smaller and eventually loses useful content. BANDMAS therefore exposes a controllable latency--utility trade-off rather than guaranteeing dominance at every load.

\textbf{Scheduling overhead remains small relative to inference.} Causal reduces estimated receiver context from 53.7 to 23.5 tokens on SciFact, 232.1 to 68.0 tokens on HotpotQA, and 229.3 to 122.5 tokens on FanOutQA. Scheduler latency reaches 6.63~ms on FanOutQA but remains below 1\% of p95 latency. Persistent Ollama allocation makes warmed memory deltas uninformative, so GPU telemetry remains an artifact rather than packet-level KV evidence.

Typed packets establish a tighter admission boundary. Replay-derived scoring distinguishes contributions, and adaptation restrains tail growth after queues build at a utility cost under the heaviest loads. This method supports one core claim: it cuts application-layer traffic and enables benchmark-specific queue control under frozen generated traffic and controlled protocol rules. Still, this approach cannot guarantee lower latency across all operating scenarios.

\subsection{Design Implications}

\textbf{Packet boundaries and packet values solve different problems.} Separate ablation studies demonstrate that typed granularity creates schedulable choices, whereas replay-derived valuation identifies which choices matter. Moderate rank correlations motivate the auditable shrinkage estimator; richer predictors must preserve outcome isolation and include their own latency and calibration cost.

\textbf{Static and adaptive admission target different regimes.} Static ranking is sufficient below saturation; completed-service pricing becomes useful after link or receiver queues build, where lower tail latency may require lower utility. Deployments should therefore select a deadline-miss or utility target rather than expect unconditional improvement. Reported byte savings apply to compact application records, not transport headers, retransmissions, encryption, or model-serving effects.

\section{Threats and Limitations}

RCE is conditional on frozen messages, receiver utility, and two replay contexts; it does not identify unrestricted adaptive-agent effects. Evaluation covers two Qwen generations, three evidence-aggregation workloads, and controlled lossless links. Baselines are protocol-matched reproductions, byte counts exclude transport headers, and invalid producer traffic remains outside scope~\cite{zheng2022poisoning,lin2021medley}.

\section{Use of AI Disclosure}

OpenAI Codex assisted with experiment planning, software tests, analysis scripts, figures, and language editing. The authors reran analyses, verified reported numbers against result files, reviewed all prose, and take responsibility for the study and manuscript.

\section{Conclusion}

LLM-based multi-agent workflows must exchange useful evidence without forwarding every generated message. BANDMAS converts messages into typed, traceable semantic packets and schedules complete records using predicted replay-derived contribution and resource pressure. It separates offline causality-inspired valuation from online admission through an auditable no-outcome-leakage boundary. Evaluation shows packetization and ranking improve communication efficiency, while adaptive pricing exposes a load-dependent latency--utility trade-off. BANDMAS provides a communication-control interface between multi-agent collaboration and network resource management.

\bibliographystyle{IEEEtran}
\bibliography{references}

\end{document}